\documentstyle[pra,preprint,aps]{revtex}
\tightenlines
\RequirePackage{epsfig}
\def\Journal#1#2#3#4{{#1} {\bf #2}, #3 (#4)}
\def\PR{Phys. Rev.}
\def\PRL{Phys. Rev. Lett.}
\def\PRA{Phys. Rev. A}
\def\JMP{J. Math. Phys.}

\def\Science{Science}
\def\EPJD{Eur. Phys. J. D}
\def\RMP{Rev. Mod. Phys.}

\def\JPB{J. Phys. B: At. Mol. Opt. Phys.}
\begin{document}
\draft
\title {Quantum Mechanics of One-Dimensional Trapped Tonks Gases}
\author{M. D. Girardeau\thanks{Email: girardeau@optics.arizona.edu}
and E. M. Wright\thanks{Email: Ewan.Wright@optics.arizona.edu}}
\address{Optical Sciences Center\\ University of Arizona\\ Tucson, AZ 85721, USA}
\date{\today}
\maketitle
\begin{abstract}
Several experimental groups are currently working towards
realizing quasi-one-dimensional (1D) atom de Broglie waveguides
and loading them with ultracold atoms. The dynamics becomes truly
1D in a regime (Tonks gas) of low temperatures and densities, and
large positive scattering lengths for which the transverse mode
becomes frozen, in which case the many-body Schr\"{o}dinger
dynamics becomes exactly soluble via a Fermi-Bose mapping theorem.
In this paper we review our recent work on the exact ground state
and quantum dynamics of 1D Tonks gases and assess the possibilty
of approaching the Tonks regime using Bessel beam optical dipole
traps.
\end{abstract}
\pagebreak
\section{Introduction}
Recent advances in atom de Broglie waveguide technology
\cite{Key,Muller,Dekker,Thy,Hinds} and its potential applicability
to atom interferometry \cite{Berman} and integrated atom optics
\cite{Dekker,Sch} create a need for accurate theoretical modelling
of such systems in the low temperature, tight waveguide regime
where transverse excitations are frozen out and the quantum
dynamics becomes essentially one-dimensional (1D) (Tonks-gas
limit). It has been shown by Olshanii \cite{Olshanii}, and also
recently by Petrov et al. \cite{PetShlWal00}, that at sufficiently
low temperatures and densities, high transverse frequencies
$\omega_T$, and large positive scattering length, where thermal
and longitudinal zero-point energies are small compared with
$\hbar\omega_T$, a Bose-Einstein condensate (BEC) in a thin
cigar-shaped trap has dynamics which reduce to those of a 1D gas
of hard core, or impenetrable, point bosons. This is a model for
which the exact many-body energy eigensolutions were found in 1960
using an exact mapping from the Hilbert space of energy
eigenstates of an {\em ideal} gas of spinless fermions to that of
many-body eigenstates of hard core, and therefore {\em strongly
interacting}, bosons \cite{map,map2}. In this limit there are
strong short-range pair correlations which are omitted in the
Gross-Pitaevskii (GP) approximation, which assumes that all $N$
bosons occupy the same orbital (complete BEC, condensed fraction
unity). In the absence of a trap potential it is known
\cite{Lenard} that the occupation of the lowest orbital is of
order $\sqrt{N}$ where $N$ is the total number of atoms, in
contrast to $N$ for the ideal Bose gas as well as the GP
approximation.  Nevertheless, this system exhibits some BEC-like
behavior such as Talbot recurrences following an optical lattice
pulse \cite{Rojo} and dark soliton-like behavior in response to a
phase-imprinting pulse \cite{soliton}.

The case of harmonically trapped, hard core bosons in 1D is more
relevant to recent atom waveguide experiments \cite{BonBurDet00}.
The spatial profile of the single-particle density is expressible
in closed form, and has recently been shown \cite{Kolomeisky} to
be well approximated by a modified 1D effective field theory,
although we have recently shown in a numerically accurate
time-dependent calculation \cite{breakdown} that spatial
interference fringes of separated and recombined condensates in
the exact many-body solution are much weaker than those predicted
by the corresponding time-dependent mean field theory
\cite{Kolomeisky}. Although the Fermi-Bose mapping theorem
\cite{map,map2} implies that all physical properties expressible
in terms of spatial configurational probabilities are the same for
the actual bosonic system and the fictitious ``spinless fermion"
system used for the mapping, the momentum distribution of the
bosonic system, or more generally its occupation distribution over
the relevant orbitals for a given geometry, is very different in
the bosonic system. It is known \cite{Olshanii,Lenard,Vaidya} that
for a spatially uniform system of hard core bosons in 1D, the
momentum distribution is strongly peaked in the neighborhood of
zero momentum, whereas that of the corresponding Fermi system is
merely a filled Fermi sea. In the case of hard core bosons in a 1D
harmonic trap, it has been an interesting  question whether the
system undergoes true BEC or merely an attenuated one such as that
in the uniform system.  Ketterle and Van Druten \cite{KetVan96}
have shown that true BEC occurs for a finite number of atoms in a
1D harmonic oscillator (HO) in the case of an {\em ideal} gas, but
the behavior turns out to be different for the exact many-body
solutions in the Tonks (1D) limit. The most fundamental definition
of BEC and the condensate orbital is based on the large distance
behavior of the one-particle reduced density matrix
$\rho_{1}(x,x')$. If off-diagonal long-range order (ODLRO) is
present and hence the largest eigenvalue of $\rho_1$ is
macroscopic (proportional to $N$) then the system is said to
exhibit true BEC and the corresponding eigenfunction, the
condensate orbital, plays the role of an order parameter
\cite{PO,Yang2}. Although the precise definition of ODLRO requires
a thermodynamic limit not strictly applicable to mesoscopic traps,
the GP approximation assumes from the start that ODLRO and
macroscopic occupation of a single orbital are good approximations
in a trap, so examination of this assumption is important. We have
recently \cite{1dsho} used the Fermi-Bose mapping theorem to
determine, for this case of 1D impenetrable bosons in a harmonic
trap, the exact many-body ground state and its salient features,
including the one-particle reduced density matrix and its
eigenvalues (occupation number distribution function) and
eigenfunctions (natural orbitals), as well as the momentum
distribution function. We find that the largest eigenvalue of the
one-particle density matrix is proportional to lower than the
first power of $N$, as is the momentum distribution function at
$k=0$, so that the Tonks gas does not show true BEC.

In this paper we give an overview of our recent work on exact
many-body quantum solutions for 1D Tonks gases. In particular,
Secs. II and III describe the ground state and dynamic properties
of Tonks gases, respectively. For the most part we investigate 1D
Tonks gases assuming the conditions are satisfied for realizing
them. In Sec. IV we examine the use of Bessel optical dipole traps
for experimental realization of 1D Tonks gases.
\section{Exact stationary solutions}
We first describe the Fermi-Bose mapping method for obtaining
exact solutions of the time-independent many-body Schr\"{o}dinger
equation in the Tonks limit (impenetrable point bosons) and its
application to the spatially uniform case (no trap potential).
This description is very brief and the original literature
\cite{map,map2} should be consulted for details. Some salient
features of the recent generalization to the harmonically trapped
Tonks gas \cite{1dsho} are then described.
\subsection{Mapping theorem}
Our basic model consists of $N$ bosonic atoms at zero temperature
moving in 1D ($x$), the quantum motion in the other two spatial
dimensions having been frozen out by tight transverse confinement
via an atomic waveguide. The Schr\"{o}dinger Hamiltonian is then
assumed to have the structure:
\begin{equation}\label{eq1}
\hat{H}=\sum_{j=1}^{N}-\frac{\hbar^2}{2m}\frac{\partial^2}{\partial x_{j}^{2}}
+V(x_{1},\cdots,x_{N})  ,
\end{equation}
where $x_j$ is the one-dimensional position of the $j{\it th}$
particle (atom) and $V$ is symmetric (invariant) under
permutations of the particles. The two-particle interaction
potential is assumed to contain a hard core of 1D diameter $a$.
This is conveniently treated as a constraint on allowed wave
functions $\psi(x_{1},\cdots,x_{N})$:
\begin{equation}\label{eq2}
\psi=0\quad\text{if}\quad |x_{j}-x_{k}|<a\quad,\quad 1\le j<k\le N  ,
\end{equation}
rather than as an infinite contribution to $V$, which then consists of all
other (finite) interactions and external potentials. Let
$\psi_{F}(x_{1},\cdots,x_{N})$ be a fermionic solution of $\hat{H}\psi=E\psi$
which is antisymmetric under
all particle pair exchanges $x_{j}\leftrightarrow x_{k}$, hence all
permutations. One can consider $\psi_F$ to be either the wave function of
a fictitious system of ``spinless fermions'', or else that of a system of
real fermions whose spins are all aligned, as in magnetically trapped
atomic vapor BECs. Define a ``unit antisymmetric function" \cite{map}
\begin{equation}\label{eq3}
A(x_{1},\cdots,x_{N})=\prod_{1\le j<k\le N}\text{sgn}(x_{k}-x_{j})  ,
\end{equation}
where $\text{sgn}(x)$ is the algebraic sign of the coordinate difference
$x=x_{k}-x_{j}$, i.e., it is +1(-1) if $x>0$($x<0$). For given
antisymmetric $\psi_F$,
define a bosonic wave function $\psi_B$ by
\begin{equation}\label{eq4}
\psi_{B}(x_{1},\cdots,x_{N})=A(x_{1},\cdots,x_{N})\psi_{F}(x_{1},\cdots,
x_{N})
\end{equation}
which defines the Fermi-Bose mapping. $\psi_B$ satisfies
the hard core constraint (2) if $\psi_F$ does, is totally
symmetric (bosonic) under permutations, obeys the same
boundary conditions as $\psi_F$, and $\hat{H}\psi_{B}=E\psi_{B}$ follows from
$\hat{H}\psi_{F}=E\psi_{F}$ \cite{map,map2}. In the case of periodic
boundary conditions (no trap potential, spatially uniform system) one must add
the proviso that the
boundary conditions are only preserved under the mapping if $N$ is odd,
but the case of even $N$ is easily accomodated by imposing periodic
boundary conditions on $\psi_F$ but {\em anti}periodic boundary conditions
on $\psi_B$. For a trapped system the boundary condition that
wave functions vanish at infinity is preserved for all $N$.
\subsection{Exact solution for untrapped bosons}
The mapping theorem leads to explicit expressions for all
many-body energy eigenstates and eigenvalues of a 1D scalar
condensate (bosons all of the same spin) under the assumption that
the only two-particle interaction is a zero-range hard core
repulsion, represented by the $a\rightarrow 0$ limit of the
hard-core constraint. Such solutions were obtained in Sec. 3 of
the original work \cite{map} for periodic boundary conditions and
no external potential. Such a situation could be realized
experimentally, for example, by trapping the Tonks gas around the
peak of an tight toroidal optical dipole trap formed using a
red-detuned Laguerre-Gaussian laser beam \cite{WriArlDho00}. In a
system of 1D bosons with a hard-sphere interaction of diameter
$a$, it is sufficient at low densities \cite{Olshanii,PetShlWal00}
to consider the case of impenetrable point particles, the
zero-range limit $a\rightarrow 0$. Since wave functions of
``spinless fermions" are antisymmetric under coordinate exchanges,
their wave functions vanish automatically whenever any
$x_{j}=x_{k}$, the constraint has no effect, and the corresponding
fermionic ground state is the ground state of the {\em ideal} gas
of fermions, a Slater determinant of the lowest $N$
single-particle plane-wave orbitals. The exact many body ground
state was found to be a pair product of Bijl-Jastrow form
\begin{equation}
\psi_{0}=\text{const.}\prod_{i>j}|\sin[\pi L^{-1}(x_{i}-x_{j})]| ,
\end{equation}
where $L$ is the perimeter of the annular trap. In spite of the
very long range of the individual pair correlation factors
$|\sin[\pi L^{-1}(x_{i}-x_{j})]|$, the pair distribution function
$D(x_{ij})$, which is physically the joint probability density
that if one particle is found at $x_i$ a second will be found at
$x_j$, was found to be of short range
\begin{equation}
D(x_{ij})=1-\left (\frac{\sin(\pi\rho x_{ij})}{\pi\rho x_{ij}}
\right )^2 ,
\end{equation}
with $\rho=N/L$ the linear number density of the system. Clearly,
$D(0)=0$ which reflects the hard core nature of the two-particle
interaction. By examination of the excited states the system was
also found to support propagation of sound with speed
$c=\pi\hbar\rho/m$ \cite{map}.
\subsection{Harmonically trapped Tonks gas}
Here we briefly review our recently-obtained exact solution for the
many-body ground state of a harmonically trapped Tonks gas \cite{1dsho}.
The Hamiltonian of N bosons in a 1D harmonic trap is
\begin{equation}
\hat{H}=\sum_{j=1}^{N}
\left[-\frac{\hbar{^2}}{2m}\frac{\partial^2}{\partial x_{j}^{2}}
+\frac{1}{2}m\omega^{2}x_{j}^{2}\right]  .
\end{equation}
We again assume that the two-body interaction potential consists only of a hard
core of negligible diameter $a\to 0$. It follows from the mapping theorem
that the exact N-boson ground state $\psi_{B0}$ is
\begin{equation}
\psi_{B0}(x_{1},\cdots,x_{N})=|\psi_{F0}(x_{1},\cdots,x_{N})|  ,
\end{equation}
where $\psi_{F0}$ is the ground state of a fictitious system of
$N$ spinless fermions with the same Hamiltonian and constraint. Since
the hard-core interaction has no effect in the zero-range limit
for fermions, whose wave functions already vanish at contact
due to antisymmetry, it follows that the fermionic ground state is a Slater
determinant of
the lowest $N$ single-particle eigenfunctions $\phi_n$ of the
harmonic oscillator (HO):
\begin{equation}
\psi_{F0}(x_{1},\cdots,x_{N})=\frac{1}{\sqrt{N!}}
\det_{(n,j)=(0,1)}^{(N-1,N)}\phi_{n}(x_{j}) .
\end{equation}
The HO orbitals are
\begin{equation}
\varphi_{n}(x)= \frac{1}
{\pi^{1/4}x_{osc}^{1/2}\sqrt{2^{n}n!}}e^{-Q^{2}/2}H_{n}(Q)
\end{equation}
with $H_n(Q)$ the Hermite polynomials and
$Q=x/x_{osc},x_{osc}=\sqrt{\hbar/m\omega}$
being the ground state width of the harmonic trap for a single atom.
By factoring the Gaussians out of the determinant and carrying out
elementary row and column operations, one can cancel all terms in each
$H_n$ except the one of highest degree, with the result \cite{Aitken}
\begin{eqnarray}
\det_{(n,j)=(0,1)}^{(N-1,N)}H_{n}(x_{j})
& = & 2^{N(N-1)/2}\det_{(n,j)=(0,1)}^{(N-1,N)}(x_{j})^{n} \nonumber\\
& = & 2^{N(N-1)/2}\prod_{1\le j<k\le N}(x_{k}-x_{j})
\end{eqnarray}
Substitution into (6) then yields a simple but exact analytical
expression
of Bijl-Jastrow pair product form for the $N$-boson ground state:
\begin{equation}
\psi_{B0}(x_{1},\cdots,x_{N})=C_{N}\left[\prod_{i=1}^{N}e^{-Q_{i}^{2}/2}
\right]
\prod_{1\le j<k\le N}|x_{k}-x_{j}|
\end{equation}
with $Q_{i}=x_{i}/x_{osc}$ and normalization constant
\begin{equation}
C_{N}=2^{N(N-1)/4}\left (\frac{1}{x_{osc}} \right )^{N/2}
\left[N!\prod_{n=0}^{N-1}n!\sqrt{\pi}\right]^{-1/2}  .
\end{equation}
It is interesting to
note the strong similarity between this exact 1D $N$-boson wave
function and the famous Laughlin variational wave function of the 2D ground
state for the quantized fractional Hall effect \cite{Laughlin},
as well as the closely-related wave functions for bosons with weak repulsive
delta-function interactions in a harmonic trap in 2D found recently
by Smith and Wilkin \cite{SW}.

Both the single particle density and pair distribution function depend
only on the absolute square of the many-body wave function, and since
$|\psi_{B0}|^{2}=|\psi_{F0}|^{2}$ they reduce to standard ideal Fermi
gas expressions. The single particle density, normalized to $N$, is
\begin{equation}
\rho(x)=N\int |\psi_{B0}(x,x_{2},\cdots,x_{N})|^{2}dx_{2}\cdots dx_{N}
=\sum_{n=0}^{N-1}|\varphi_{n}(x)|^{2}
\end{equation}
We shall not exhibit it here since it has recently been calculated by
Kolomeisky {\it et al.} \cite{Kolomeisky}; see also our recent discussion of
the time-dependent case \cite{breakdown}.
The pair distribution function, normalized to $N(N-1)$, is
\begin{eqnarray}
& & D(x_{1},x_{2})=N(N-1)\int
|\psi_{B0}(x_{1},\cdots,x_{N})|^{2}dx_{3}
\cdots dx_{N} \nonumber\\
& & =\sum_{0\le n<n'\le N-1}\hspace{-0.5cm}
|\varphi_{n}(x_{1})\varphi_{n'}(x_{2})
-\varphi_{n}(x_{2})\varphi_{n'}(x_{1})|^{2}
\end{eqnarray}
Noting that terms with $n=n'$, which vanish by antisymmetry, can be
formally added to the summation (13), one can write
\begin{eqnarray}
D(x_{1},x_{2}) & = & \rho(x_{1})\rho(x_{2})
-|\Delta(x_{1},x_{2})|^{2} \nonumber\\
\Delta(x_{1},x_{2}) & = &
\sum_{n=0}^{N-1}\varphi_{n}^{*}(x_{1})\varphi_{n}(x_{2})
\end{eqnarray}
Although the Hermite polynomials have disappeared from
the expression (10) for the many-body wave function, they reappear upon
integrating $|\psi_{B0}|^{2}$ over $(N-1)$ coordinates to get the single
particle density $\rho(x)$ and over $(N-2)$ to get the pair distribution
function $D(x_{1},x_{2})$, and the expressions in terms of the HO orbitals
$\varphi_n$ are the most convenient for evaluation.
Figure \ref{Fig:one}
shows a gray scale plot of the dimensionless pair distribution function
$x_{osc}^2\cdot D(Q_1,Q_2)$ versus the normalized coordinates
$Q_{1,2}=x_{1,2}/x_{osc}$
for a) $N=2$, b) $N=6$, and c) $N=10$.  Some qualitative features
of the pair distribution function are apparent: In the first place it
follows either from the original expression (12) or from Eq. (14)
that $D(x_1,x_2)$ vanishes at contact $x_1=x_2$, as it must because of
impenetrability of the particles, and we see this to be true in Fig.
\ref{Fig:one}.  Furthermore, the correlation term
$\Delta(x_{1},x_{2})$ is a truncated closure sum and approaches the
Dirac delta function $\delta(x_{1}-x_{2})$ as $N\rightarrow\infty$, as is to
be expected since the healing length in a spatially uniform 1D hard core
Bose gas varies inversely with particle number \cite{soliton}.
As a result the width of the null around the diagonal $Q_1=Q_2$
decreases with increasing $N$, and vanishes in the limit.  Away from
the diagonal along $Q_2=-Q_1$ the pair distribution function rises,
exhibits modulations for $N>2$, due to the oscillatory nature of the
HO orbitals, before decreasing back to zero at large distances.
For $|x_{1}-x_{2}|$ much larger than
the healing length, $D$ reduces to the uncorrelated density product
$\rho(x_{1})\rho(x_{2})$, so the spatial extent of the pair distribution
function is that of the density and varies as $N^{1/2}$ \cite{Kolomeisky}.

The reduced single-particle density matrix with
normalization $\int\rho_{1}(x,x)dx=N$ is
\begin{equation}
\rho_{1}(x,x')=N\int\psi_{B0}(x,x_{2},\cdots,x_{N})
\psi_{B0}(x',x_{2},\cdots,x_{N})dx_{2}\cdots dx_{N}
\end{equation}
Although this multi-dimensional integral cannot be evaluated
analytically, it can be evaluated numerically by Monte Carlo integration
for not too large values of $N$ (the computing time scales as $N^4$).
Figure \ref{Fig:two}
shows a gray scale plot of the dimensionless reduced single-particle
density matrix
$x_{osc}\cdot\rho_1(Q,Q')$ versus the normalized coordinates
$Q$ and $Q'$ for a) $N=2$, b) $N=6$, and c) $N=10$.
Along the diagonal $\rho_1(Q,Q'=Q)=\rho(Q)$ reproduces the single-particle
density \cite{Kolomeisky}.  The off-diagonal elements relate to ODLRO, and it
is clear that as $N$ increases the off-diagonal elements are decreasing, in
contrast with the diagonal.  This is an indication that ODLRO vanishes
for a system of hard core bosons in a 1D HO in the thermodynamic
limit, although at present only numerical evidence exists, there being
no analytical proof generalizing the result of Lenard for the untrapped
Tonks gas \cite{Lenard}.

In a macroscopic system, the presence or absence of BEC is determined by the
behavior of $\rho_{1}(x,x')$ as $|x-x'|\rightarrow\infty$. Off-diagonal
long-range order is present if the largest eigenvalue of
$\rho_1$ is macroscopic (proportional to $N$), in which case the system
exhibits BEC and the corresponding eigenfunction, the condensate
orbital, plays the role of an order parameter \cite{PO,Yang2}. Although
this criterion is not strictly applicable to mesoscopic systems, if the
largest eigenvalue of $\rho_1$ is much larger than one
then it is reasonable to expect that
the system will exhibit some BEC-like coherence effects. Thus we examine here
the spectrum of eigenvalues $\lambda_j$ and associated eigenfunctions
$\phi_{j}(x)$ (``natural orbitals'') of $\rho_1$. Although natural orbitals
are a much-used tool in theoretical chemistry, they have only
recently been applied to mesoscopic atomic condensates \cite{DuBGly00}.
The relevant eigensystem equation is
%
\begin{equation}
\int_{-\infty}^{\infty}\rho_{1}(x,x')\phi_{j}(x')dx'=\lambda_{j}\phi_{j}(x)
\end{equation}
$\lambda_j$ represents the occupation of the orbital $\phi_j$, and
one has $\sum_{j}\lambda_{j}=N$. Numerical evaluation of the
integral by discretization yields a matrix eigensystem equation
giving accurate numerical results for the largest eigenvalues and
associated eigenvectors.  We remark that for the corresponding
problem of $N$ free fermions in a 1D HO, in which case
$\rho_1(x,x')$ is evaluated using the fermion wave function
$\psi_{F0}(x_{1},\cdots,x_{N})$, the natural orbitals are simply
the HO orbitals, and $\lambda_j=1, j=0\cdots (N-1)$, all higher
eigenvalues being zero (filled Fermi sea).  However, the N-boson
wave function is the modulus of the fermion wave function, and
this leads to significant differences in the spectrum of natural
orbitals and eigenvalues for the hard core Bose gas. In Fig.
\ref{Fig:three}(a) we show a log-log plot of the fractional
occupation of the lowest orbital $f_0=\lambda_0/N$ versus the
total particle number $N$ (solid line), along with a best fit
power-law $f_0\approx N^{-0.41}$ (dashed line). This is to be
contrasted with the case of a spatially uniform system of hard
core bosons for which $f_0\approx N^{-0.5}$ \cite{Lenard}. In both
cases the fractional occupation decreases with increasing $N$, and
thus does not correspond to a true condensate for which $f_0=1$.
Nevertheless, the occupation of the lowest orbital may still be
large, $\lambda_0\approx N^{0.59}$, and is larger than the
spatially uniform case $\lambda_0\approx N^{0.5}$, so macroscopic
quantum coherence effects reminiscent of BEC can still result
\cite{Olshanii,Lenard,Rojo,soliton,Kolomeisky,breakdown}. Figure
\ref{Fig:three}(b) shows the distribution of occupations
$\lambda_j$ versus orbital number $j$ (the orbitals are ordered
according to eigenvalue magnitude, the largest eigenvalue being
$j=0$) for $N=2$ (circles), $N=6$ (stars), and $N=10$ (squares).
This figure shows that as the lowest orbital occupation
$\lambda_0$ increases with increasing $N$ so does the range of
significantly occupied higher-order orbitals with $j>0$.  This
means that the dominance of the lowest orbital decreases with
increasing $N$, so singling out $\phi_0(x)$ as a macroscopic wave
function for the whole system becomes more problematic with
increasing $N$ \cite{Kolomeisky,breakdown}.

Next we examine the momentum distribution for the trapped 1D Tonks
gas. For a spatially uniform system (no trap) the natural orbitals
are plane waves, so the occupation distribution of the natural
orbitals is the same as the momentum distribution. Although this
is not the case here due to the effect of the harmonic trap
potential, the momentum distribution is still physically
important. In terms of the boson annihilation and creation
operators in position representation (quantized Bose field
operators) the one-particle reduced density matrix is
\begin{equation}
\rho_{1}(x,x')=\langle\Psi_{B0}|\hat{\psi}^{\dagger}(x')\hat{\psi}(x)|
\Psi_{B0}\rangle
\end{equation}
The momentum distribution function $n(k)$, normalized to
$\int_{-\infty}^{\infty} n(k)dk=N$, is
$n(k)=\langle\Psi_{B0}|\hat{a}^{\dagger}(k)\hat{a}(k)|\Psi_{B0}\rangle$
where $\hat{a}(k)$ is the annihilation operator for a boson with
momentum $\hbar k$. Then
\begin{equation}
n(k)=(2\pi)^{-1}\int_{-\infty}^{\infty}dx\int_{-\infty}^{\infty}dx'
\rho_{1}(x,x')e^{-ik(x-x')}
\end{equation}
The spectral representation of the density matrix then leads to
$n(k)=\sum_{j}\lambda_{j}|\mu_{j}(k)|^2$ where the $\mu_j$ are
Fourier transforms of the natural orbitals:
$\mu_{j}(k)=(2\pi)^{-1/2}\int_{-\infty}^{\infty}\phi_{n}(x)e^{-ikx}dx$.
Figure \ref{Fig:four} shows the numerically calculated
dimensionless momentum spectrum $k_{osc}\cdot n(\kappa)$ versus
normalized momentum $\kappa=k/k_{osc}$, with
$k_{osc}=2\pi/x_{osc}$, for  a) $N=2$, b) $N=6$, and c) $N=10$.
The key features are that the momentum spectrum maintains the
sharp peaked structure reminiscent of the spatially uniform case
\cite{Olshanii,Lenard} for the 1D HO, and that the peak becomes
sharper with increasing atom number $N$.  This is to be expected
since as the number of atoms increase the many-body repulsion
causes the system to become more spatially uniform within the trap
interior.

By way of contrast, for a 1D Fermi gas the corresponding momentum
spectrum is a filled Fermi-sea, here rounded out by the trapping
potential. In particular, the momentum distribution for the 1D
trapped Fermi gas can be expressed as
\begin{equation}
n(k)=\sum_{j=1}^N|\mu_{j}(k)|^2 ,
\end{equation}
Thus, the momentum spectrum provides a means of distinguishing
between the 1D Fermi and Tonks gases. Figure \ref{Fig:five} shows
an example of the Tonks (dashed line) and Fermi gas (solid line)
momentum spectra for $N=10$ atoms in a harmonic trap, and the
difference is clearly seen. Furthermore, the distinction only
grows larger with increasing number of atoms, as the filled
Fermi-sea broadens whereas the Tonks spectrum becomes narrower. In
a recent paper we have devised a scheme to measure the momentum
spectrum of trapped gases based on Raman outcoupling, and we refer
the reader to our paper for details \cite{momentum}.
\section{Dynamical solutions}
The Fermi-Bose mapping theorem is very easily generalized so as to yield
exact solutions of the time-dependent many-body Schr\"{o}dinger equation
(TDMBSE) by noting that since the mapping function $A(x_{1},\cdots x_{n})$
of Eq. (3) is
independent of time, one can merely replace $E\psi$ by
$i\hbar\partial\psi/\partial t$ ,
implying that the relationship of Eq. (4), with time arguments added
to the Bose and Fermi wave functions, is valid for solutions of the TDMBSE.
In the special case where the only interatomic interaction is that of
hard cores of vanishing diameter, but external potentials including a
trap potential as well as time-dependent fields may be present, the
many-fermion solutions of the TDMBSE are Slater determinants of solutions
of the {\em single particle} TDSE in the given external potential,
and each many-boson solution is obtained by multiplying the corresponding
determinant by the mapping function $A(x_{1},\cdots x_{n})$. Some salient
features of our recently-obtained dynamical solutions for solitons in a ring
geometry and for interference effects in a dynamically split,
harmonically trapped Tonks gas will be described. The description will
again be brief since details are available in the literature
\cite{soliton,breakdown}.
\subsection{Dark solitons in a Tonks gas}
Dark and gray solitons are a generic feature of the nonlinear
Schr\"{o}dinger equation with repulsive interactions, and several
calculations of their dynamics
based on the mean-field Gross-Pitaevskii (GP) equation have appeared
\cite{Reinhardt,Dum,Scott,Jackson,Burger,Denschlag,Muryshev,Busch},
as well as experiments demonstrating their existence in atomic BECs
\cite{Burger,Denschlag}.
Since the underlying {\it many-body} Schr\"{o}dinger equation is linear,
this raises the question of how observed solitonic behavior arises.
Here this issue will be examined with the aid of
exact many-body solutions for the Tonks gas.
The model consists of a 1D hard-core Bose gas
in a toroidal trap, or ring, with cross section so small
that motion is essentially circumferencial. The Fermi-Bose
mapping is employed to generate exact solutions for this problem.
We identify
stationary solutions which reflect some properties of dark solitons
from the GP theory when the ring is pierced at a point by an intense
blue-detuned laser.  We also present dynamical solutions when half of
an initially homogeneous ring BEC is phase-imprinted via the light-shift
potential of an applied laser, leading to gray soliton-like
structures whose velocity depends on the imposed phase-shift
\cite{Burger,Denschlag}.  Such structures are apparent for times less than
the echo time $\tau_e=L/c$, with $L$ the ring circumference and $c$ the
speed of sound in the BEC.  On longer time scales the dynamics becomes
very complex showing Talbot recurrences which are beyond
the GP theory.

Consider $N$ bosons in a tight toroidal trap, for example a
toroidal optical dipole potential \cite{WriArlDho00}, and denote
their 1D positions measured around the circumference by $x_j$.
This is equivalent to the exactly-solved model \cite{map} of $N$
impenetrable point bosons in 1D with wave functions satisfying
periodic boundary conditions with period $L$ equal to the torus
circumference, and the fundamental periodicity cell may be chosen
as $-L/2<x_{j}<L/2$. However, the rotationally invariant quantum
states of this problem do not reveal any dark soliton-like
structures.  To proceed we therefore consider the case that a
blue-detuned laser field pierces the ring at $x=0$ by virtue of
the associated repulsive dipole force: The light sheet then
provides a reference position for the null of the dark soliton.
Assume that the light sheet is so intense and narrow that it may
be replaced by a constraint that the many-body wave function
(hence the orbitals $\phi_i$) must vanish whenever any $x_{j}=0$.
Then the appropriate orbitals $\phi_{i}(x)$ are free-particle
energy eigenstates vanishing at $x=0$ and periodic with period
$L$. The complete orthonormal set of even-parity eigenstates
$\phi_{n}^{(+)}$ and odd-parity eigenstates $\phi_{n}^{(-)}$ are
\begin{eqnarray}\label{lightsheet}
\phi_{n}^{(+)}(x) & = & \sqrt{2/L}\sin[(2n-1)\pi|x|/L]  , \nonumber\\
\phi_{n}^{(-)}(x) & = & \sqrt{2/L}\sin(2n\pi x/L)       ,
\end{eqnarray}
with $n$ running from $1$ to $\infty$. The odd eigenstates are
the same as those of free particles with no $x=0$ constraint, since these
already vanish at $x=0$. However, the even ones are strongly affected by the
constraint, their cusp at $x=0$ being a result of the impenetrable light sheet
at that point. If one bends a 1D box $-L/2<x<L/2$ with impenetrable
walls into a ring, identifying the walls at $\pm L/2$, then those
particle-in-a-box eigenfunctions which are even about the box center become
identical with the $\phi_{n}^{(+)}$, and their cusp results from
the nonzero slope of these functions at the walls. The $N$-fermion ground state
is obtained by inserting the lowest $N$ of these orbitals into a Slater
determinant (filled Fermi sea).  Since $A^{2}=1$, the one-particle
density $\rho(x)$ of the corresponding many-boson ground state given by
Eq. (4) is the same as that of the $N$-fermion ground state. In the
thermodynamic limit $N\rightarrow\infty$, $L\rightarrow\infty$,
$N/L\rightarrow\rho$ for fixed
$x$, $\rho^{(\pm)}$, one finds \cite{soliton}
\begin{equation}
\rho(x)\sim \rho[1-\sin(2\pi\rho x)/2\pi\rho]  .
\end{equation}
$\rho(x)$ vanishes at $x=0$ and approaches the mean density
$\rho$ over a healing length $L_h=1/2\rho$ with damped spatial
oscillations about its limiting value. Suppose next that the
light-sheet is turned off at $t=0$ by removing the constraint that
the wave function vanish at $x=0$. The solution of the TDMBSB for
the many-boson system is then still given by the mapping theorem,
but the Slater determinant representing the corresponding
many-fermion state has to be built from solutions of the
time-dependent single-particle Schr\"{o}dinger equation satisfying
the initial conditions of Eq. (\ref{lightsheet}). The odd-parity
solutions are stationary in time since they already vanish at the
position of the light sheet. The even-parity solutions are
nontrivial, but are expressible as sums over the space-periodic,
even-parity solutions of the time-independent free-particle
Schr\"{o}dinger equation \cite{soliton}:
\begin{equation}
\phi_{n}^{(+)}(x,t)=\frac{2(2n-1)}{\pi}\sqrt{\frac{2}{L}}
\sum_{p=0}^{\infty}\frac{(2-\delta_{p0})\cos(k_{p}x)e^{-i\omega_{p}t}}
{(2n-1)^{2}-4p^{2}}
\end{equation}
where $\omega_{p}=\hbar k_{p}^{2}/2m$ and $k_{p}=2p\pi/L$.
The time-dependent density $\rho(x,t)$ is then the sum of absolute squares
of all $N$ orbitals in the Fermi sea. It is found \cite{soliton} that
there are two important time scales: One is the Poincar\'{e}
recurrence time $\tau_r$. Noting that $\omega_p$ is proportional to
$p^2$, one finds that all terms in the sum are time-periodic with period
$\tau_{r}=mL^{2}/\pi\hbar$, which is therefore the recurrence time for the
density and in fact all properties of our model \cite{Rojo}.
The other important time is the echo
time $\tau_e$, the time for sound to make one circuit around the torus.
Recalling
that the speed of sound in this system is $c=\pi\hbar\rho/m$ \cite{map},
one finds $\tau_{e}=\tau_{r}/N$.  For $t<<\tau_e$ after the constraint is
removed, the initial density develops sound waves that propagate around
the ring, and that we examine below in the context of phase-imprinting.
For $t>\tau_e$ the evolution is very complex, but complete recurrences
occur for times $t=n\tau_r$ with fractional revivals in between.

Suppose next that the toroidal impenetrable point Bose gas is in its ground
state for times $t<0$ (no light sheet obstacle in this case), and a
phase-imprinting laser pulse is applied as a delta-function pulse over half
the ring at $t=0$. This is described by the Hamiltonian
\begin{equation}
\hat{H}=\sum_{j=1}^{N}\left[-\frac{\hbar^2}{2m}\frac{\partial^2}
{\partial x_{j}^{2}}-\hbar\Delta\theta\delta(t)S(x_{j})\right]
\end{equation}
where $S(x)=\theta(L/4-|x|)$, i.e., unity for $-L/4<x<L/4$ and
zero elsewhere. This is the technique used in recent experiments
\cite{Burger,Denschlag}, here idealized to
a delta-function in time and to sharp spatial edges. Before the pulse the
most convenient free-particle orbitals in (5)
are plane waves $\phi_{n}(x)=\sqrt{(1/L)}e^{ik_{n}x}$ where $k_{n}=2n\pi/L$
and $n=-n_{F},-n_{F}+1,\cdots,n_{F}-1,n_{F}$ with $n_{F}=(N-1)/2$. Let
$\phi_{n}(x,t)$ be the solution of the corresponding single-particle
TDSE reducing to the above
$\phi_{n}(x)$ just before the pulse. Then the solutions just
after the pulse are $\phi_{n}(x,0+)=\phi_{n}(x)e^{iS(x)\Delta\theta}$.  The
potential gradients at the pulse edges impart momentum kicks to the particles
there which induce both compressional waves propagating at the speed, $c$,
of sound and density dips (gray solitons) moving at speeds $|v|<c$.
The expansion of $\phi_{n}(x,t)$ in terms of the unperturbed plane waves is
evaluated as
\begin{eqnarray}
\phi_{n}(x,t) & = & \frac{1}{2}\left(1+e^{i\Delta\theta}\right)
-\frac{1-e^{i\Delta\theta}}{\pi}
\sum_{\ell=-\infty}^{\infty}\nonumber\\
& \times & \frac{(-1)^{\ell}\phi_{n-2\ell-1}(x)
e^{-i\omega_{n-2\ell-1}t}}{2\ell+1}
\end{eqnarray}
and the time-dependent density is the sum of the absolute squares
of the lowest $N$ of these. Figure \ref{Fig:six} shows numerical
simulations for $N=51$, $t/\tau_e=0.051$, and $\Delta\theta=\pi$
(solid line), and $\Delta\theta=0.5\pi$ (dashed line): due to
symmetry we show only half of the ring $-L/2<x<0$, the phase shift
being imposed at $x=-L/4$. Considering times short compared to the
echo time means that the corresponding results are not very
sensitive to the periodic boundary conditions, and also therefore
apply to a linear geometry. The initial density profile is flat
with a value $\rho_0 L=51$. For both phase shifts two distinct
maxima are seen, which travel at close to the speed of sound $c$,
and two distinct minima, which are analogous to gray solitons and
travel at velocities $|v|/c<1$. In addition, there are also high
wavevector oscillations which radiate at velocities greater than
$c$, analogous to precursors in electromagnetic wave propagation
in a medium. In the case of a phase shift $\Delta\theta=\pi$, the
density is symmetric about $x=-L/4$, whereas for a phase-shift
other than a multiple of $\pi$ the evolution is not symmetric, see
the dashed line where the global minimum moves to the right in
reponse to the phase shift. In Fig. \ref{Fig:seven} we plot the
calculated velocity of the global density minimum relative to the
speed of sound for a variety of phase shifts $\Delta\theta$.  The
basic trend is that larger phase-shift means lower velocity, in
qualitative agreement with recent experiments
\cite{Burger,Denschlag}, but there is a sharp velocity peak at
$\Delta\theta\approx 0.83\pi$: This peak results from the
cross-over between two local minima in the density.  These general
features, the generation of gray solitons and density waves, agree
with those of the GP theory, but here arise out of the exact
many-body calculation.

A detailed comparison between our results and current experiments is not
possible as they do not conform to the conditions for a 1D system.
However, some estimates are in order to set the
appropriate time scales: If we consider $^{87}$Rb with a ring of
circumference $L=100$ $\mu$m, and a high transverse trapping frequency
$\omega_\perp=2\pi\times 10^5$ Hz, then we are limited to atom
numbers $N<300$ \cite{Olshanii}, so these are small condensates.
We then obtain $\tau_r=4.6$ s, and $\tau_e=90$ ms for $N=51$.
Finally, we remark that since our approach relied on the mapping
between the strongly-interacting Bose system and a non-interacting
``spinless Fermi gas" model, this suggests that dark and gray solitons
should also manifest themselves in the density for the 1D Fermi system.
Although real fermions have spin,
the interactions used here to generate solitons were spin-independent.
\subsection{Supression of interference in a dynamically split and recombined
Tonks gas}
Mean-field theory (MFT) has proven remarkably successful at predicting both
the static and dynamic behavior of Bose-Einstein condensates (BECs)
in weakly-interacting atomic vapors \cite{DalGioPit99},
including the ground state properties \cite{BayPet96,KagShlWal96}, the
spectrum of collective excitations \cite{EdwDodCla96,Str96},
four-wave mixing \cite{GolPlaMey95,DenHagWen99},
matter-wave solitons \cite{MorBalBur97,ReiCla97},
and interference between BECs
\cite{NarWalSch96,JavWil97,LegSol98,WriWonCol97}.
The basic notion underlying MFT
is of a macroscopic wave function \cite{PO,Yang2}, or
order parameter, which defines the spatial mode into which a
significant fraction of the atoms condense below the critical temperature.
The macroscopic wave function typically obeys a nonlinear Schr\"odinger
equation (NLSE), the Gross-Pitaevskii equation, and is most suitable
for dilute Bose gases.  However, the success of MFT is
not assured in all cases. For example, in one-dimension (1D) a spatially
homogeneous ideal gas in its many-body ground state exhibits complete
BEC into the lowest single-particle state,
but no BEC at any nonzero temperature. Furthermore, previous exact analysis
\cite{map,map2} of a spatially uniform Tonks gas by one
of us (MG) and its extension by Lenard \cite{Lenard} and Vaidya and Tracey
\cite{Vaidya} have shown that in the many-body ground state the
occupation of the lowest single-particle state is of order $\sqrt{N}$ where
$N$ is the total number of atoms, in contrast to $N$ for usual BEC, and
similar behavior, but with a slightly larger exponent, occurs in the
harmonically trapped Tonks gas \cite{1dsho}. We have recently
shown that mean-field theory breaks down
in the analysis of the dynamics of 1D atom clouds, in that it
predicts interference effects that are absent in the exact theory
\cite{CCint}.
Kolomeisky {\it et al.} \cite{Kolomeisky}
have proposed a nonlinear Schr\"{o}dinger equation (NLSE) with
a quartic nonlinearity to extend the usual mean-field
theory for 1D atom clouds.  For a harmonic trap the ground-state
density profiles from their theory show excellent agreement with the
exact many-body results (see Fig. 1 of their paper).  The key question, then,
is whether this extended NLSE can be used in all circumstances.
To address this issue we examine the problem of a 1D atomic cloud in
the ground state of a harmonic trap that is split by a blue-detuned
laser and recombined, both using an exact many-body
treatment based on the Fermi-Bose mapping and the approximate NLSE:
the NLSE predicts interference whereas the exact analysis does not.

Consider a Tonks gas which is initially in its $N$-body ground state
in a harmonic trap \cite{1dsho}. A central Gaussian repulsive potential
simulating a blue-detuned laser field is turned on quasi-adiabatically
at time $t=0$ to split the initial state. After some time $t_{pot}$
both the harmonic trap and repulsive potential are turned off and the
two split components allowed to recombine: this is an interference
experiment of the cool, cut, interfere variety previously discussed
\cite{JavWil97}. The external potential is taken of the specific form
\begin{equation}\label{eq7}
V(x,t) = \frac{1}{2}m\omega^2x^2 + V_B\sin(\frac{\pi t}{2t_{pot}})
e^{-x^2/w^2}  ,
\end{equation}
and $V=0$ for $t>t_{pot}$.  Figure \ref{Fig:eight} shows an
illustrative example for $N=10$ with a repulsive potential of
height $V_B=20\hbar\omega$, and width $w=3x_0$, with
$x_0=\sqrt{\hbar/2m\omega}$ the ground state harmonic oscillator
width, and $\omega t_{pot}=3$. This figure shows a gray scale plot
of the single-particle density
$\rho_N(x,t)=\sum_{i=1}^N|\phi_i(x,t)|^2$ of the $N$-boson system
as functions of $\omega t$ (horizonatal axis) and position $x/x_0$
(vertical axis), with white being the highest density. The
potential height was chosen such that
$V_B>\mu=\hbar\omega(N-1/2)$, where $\mu$ is the chemical
potential of the $N$-particle oscillator ground state
\cite{Kolomeisky}, noting that the top of the $N$-particle Fermi
sea is at $n=N-1$. As expected, as the repulsive potential turns
on it splits the initial ground state into two separated
components.  Upon release at $t=t_{pot}$ the two components expand
and subsequently recombine.  What is noteworthy is that although
there is some modulation upon recombination there are no strong
interference fringes indicative of the interference provisionally
expected for bosons: this was a generic finding from our
simulations irrespective of the time scale on which the repulsive
potential was turned on \cite{JavWil97,LegSol98}.  In contrast,
the density $|\phi_i(x,t)|^2$ for each individual orbital
$i=1,\ldots,N$ can show strong fringes, but with a different
period in each case.  Thus, the minimal inteference is a result of
washing out of the individual interferences by summing over N
orbitals.  Thus, the remnant of any interference fringes decreases
with increasing $N$ and vanishes in the thermodynamic limit.
Physically, it makes sense that the interference fringes are all
but absent since the Fermi-Bose mapping shows that in this 1D
limit the system of bosons acts effectively like a system of free
fermions insofar as effects expressible only in terms of
$|\psi_{B}|^2$ are concerned, so interference is not expected
\cite{CahGla99}.  The lack of interference is therefore a
signature of the Fermi-Bose duality that occurs in 1D systems of
impenetrable particles \cite{map}.

We next turn to the mean-field description proposed by Kolomeisky
{\it et al.} \cite{Kolomeisky} for low-dimensional systems.  In
particular, they introduce an order parameter $\Phi(x,t)$, normalized
to the number of particles $N$, for such
systems.
Using energy functional arguments they deduce the following
NLSE with quartic nonlinearity for a 1D system of impenetrable
bosons:
\begin{equation}\label{eq8}
i\hbar\frac{\partial\Phi}{\partial t} =
\left [-\frac{\hbar^2}{2m}\frac{\partial^2}
{\partial x^{2}}
+V(x,t) + \frac{(\pi\hbar)^2}{2m}|\Phi|^4 \right ]\Phi  .
\end{equation}
Our goal is to compare the predictions of this NLSE for the same
cool, cut, interfere simulation in Figure \ref{Fig:eight}, with
the initial condition $\Phi(x,0)=\sqrt{\rho_N(x,0)}$ corresponding
to the exact many-body solution, all other parameters being equal.
Figure \ref{Fig:nine} shows the corresponding gray-scale plot of
the density $\rho(x,t)=|\Phi(x,t)|^2$, and two features are
apparent: First, during the splitting phase when the repulsive
potential is on there is very good overall agreement between the
exact many-body theory and the NLSE prediction.  Second, when the
split components are released and recombine they produce
pronounced interference fringes, in contrast to the exact theory.
Indeed, this interference in the MFT is to be expected on the
basis of previous theoretical work \cite{NarWalSch96}, even though
a quartic (rather than quadratic) nonlinearity is employed here.
Thus the MFT cannot accurately capture the time-dependent dynamics
in all situations.
\section{Bessel optical dipole traps for Tonks gases}
So far we have assumed that conditions are satisfied for the Tonks
limit. However, previous analysis by Olshanii \cite{Olshanii} and
Petrov et al. \cite{PetShlWal00} show that there are stringent
requirements on the temperature, transverse confinement, and
linear atom number density to approach the Tonks limit. Clearly,
what is needed first of all it to create a thin cigar shaped
atomic trap so that the transverse mode becomes frozen and the
atomic motion becomes essentially 1D. Recent experimental
developments suggest that Tonks gases may be realizable in
magnetic atom waveguides \cite{Key,Dekker,Thy}, and Bongs et al.
\cite{BonBurDet00} have proposed a hybrid trap composed of optical
dipole trap formed with a first-order LG beam combined with
magnetic longitudinal trapping. Here we examine the use of Bessel
optical dipole traps as an all-optical means to realize 1D Tonks
gases.

Ideal Bessel beams are solutions of the free-space wave equation
which propagate with unchanging beam profile along the propagation
axis which we take as $x$ in cylindrical coordinates
$(r,\theta,x)$. The electric field of a monochromatic, linearly
polarized ideal Bessel beam of order $\ell$ and frequency
$\omega_L$ is \cite{Dur}
\begin{equation}
{\bf E}(r,\theta,x,t)=\frac{{\bf \epsilon}}{2} \left (E_0
J_\ell(k_r r )e^{i(k_xx+\ell\theta-\omega_L t)} + c.c. \right ) ,
\label{Jell}
\end{equation}
where $E_0$ is a scale electric field value, $J_\ell$ is the
$\ell$th order Bessel function, $\ell>0$ is the azimuthal mode
number which we take as positive for simplicity in notation, and
$k_r$ and $k_x$ are the radial and longitudinal wavevectors such
that $k^2=k_r^2+k_x^2$ with $k=\omega_L/c=2\pi/\lambda_L$.  The
zeroth-order solution $J_0$ has a central maximum surrounded by
concentric rings of roughly equal power while the higher-order
solutions $J_\ell$ have zero on-axis intensity also with
concentric rings.

The ultility of Bessel beams for optical dipole traps lies in the
fact that they can produce very elongated, and hence low aspect
ratio quasi-1D traps. In practice, of course, one cannot produce
an ideal Bessel optical beams as it carries infinite power.
However, using axiconal optics finite power approximations to
Bessel beams can be produced which can propagate over significant
distances. The basic scheme is illustrated in Figure \ref{Fig:ten}
where an incident Laguerre-Gaussian beam of azimuthal mode index
$\ell$ is transformed into a Bessel beam of the same order. The
interested reader is referred to our recent paper
\cite{ArlDhoSonWri01} for details, but for an incident Gaussian
beam on the axiconal optics the resulting zero-order Bessel beam
intensity profile past the axicon is
\begin{equation}
I_0(r,x) \approx 2\pi k_r w_0 \left ( \frac{P_0}{\pi w_0^2/2}
\right ) \left ( \frac{x}{x_{max}} \right )\exp(-2x^2/x_{max}^2)
\, J_0^2(k_r r)  , \label{J0app}
\end{equation}
where $w_0$ is the input spot size, $P_0$ the input power,
$k_r=k(n-1)\gamma$, $n$ and $\gamma$ are the index and opening
angle of the axicon, and $x_{max}=kw_0/k_r$ which gives the
longitudinal extent of the Bessel beam. This intensity profile of
the Bessel optical beam is readily converted to an
effective dipole potential
\begin{eqnarray}
V(r,x) &=& \frac{\hbar\Gamma^2}{8\Delta}\left (\frac{I(r,x)}
{I_{Sat}} \right ) \nonumber \\ &=& \frac{1}{2}M\Omega_{r0}^2\left
(r^2+\lambda^2(x-x_{peak})^2 \right ) ,
\end{eqnarray}
with $\Delta=\omega_L-\omega_A$ the laser detuning from the
optical transition frequency $\omega_A$, $\Gamma$ the natural
linewidth of the optical transition, $I_{Sat}$ is the resonant
saturation intensity, and $I({\bf r},t)=\frac{1}{2}\epsilon_0
c|E(r,z)|^2$. For a red-detuned laser the potential is negative
and the atoms are attracted to the regions of high intensity,
whereas for a blue-detuned laser the atoms are repelled into the
low field regions. Here for a red-detuned laser we have
approximated the Bessel optical dipole potential by a parabolic
potential near the axis under conditions of tight confinement,
where $x_{peak}=x_{max}/2$ is the longitudinal position of the
peak of the Bessel beam, and
\begin{equation}
\Omega_{r0}^2 =  \exp(-1/2) \frac{\hbar\Gamma^2}{4
|\Delta|}\frac{P_0}{M I_{Sat}} \frac{k}{x_{max}}k_r^2 , \qquad
\lambda = \frac{2 \sqrt{2}}{k w_0} = \frac{2.83}{k_rx_{max}}
.\label{lambda}
\end{equation}
A red-detuned ($\Delta <0$) $J_0$ optical dipole potential
therefore provides confinement in both the radial and longitudinal
directions. Here $\Omega_{r0}$ is the radial oscillation frequency
with corresponding ground state oscillator width
$w_{r0}=\sqrt{\hbar/M\Omega_{r0}}$, and $\lambda$ is the ratio
between the longitudinal and radial trap frequencies
$\Omega_{x0}/\Omega_{r0}=\lambda$
\cite{BayPet96,PerMicHer98,KivAle99}, which also determines the
aspect ratio between the radial and longitudinal ground state
widths $w_{r0}/w_{x0}=\sqrt{\lambda}$ (in the absence of many-body
repulsion).

Petrov {\it et al.} \cite{PetShlWal00} have theoretically studied
the diagram of state for a one-dimensional gas of trapped bosons,
assuming $\lambda\ll 1$, and found that a true BEC, or at least a
quasi-condensate, with concomitant macroscopic occupation of a
single state, is only attained for high enough particle numbers
$N>N_*$ with
\begin{equation}
N_* = \left(\frac{M g w_{x0}}{\hbar^2}\right)^2 = \left(2
\left(\frac{a}{w_{r0}}\right)
\left(\frac{w_{x0}}{w_{r0}}\right)\right)^2 . \label{Nstar}
\end{equation}
For $N<N_*$ and temperatures $T < N\hbar\Omega_{x0}$, one obtains
a Tonks gas of impenetrable bosons for which hard core repulsion
between the bosonic atoms prevents them from penetrating through
each other in the one-dimensional system, and the system acquires
properties reminiscent of a one-dimensional system of fermionic
atoms. A highly elongated Bessel beam discussed, say with
parameters $\lambda_L = 1064$ nm, $P_0 = 5$ W, $x_{max} = 10$ cm,
$1/k_r=1.25$ $\mu$m, would be an ideal candidate for the
experimental realization of a Tonks gas. The low aspect ratio
$\lambda = 3.5 \times 10^{-4}$ and tight radial confinement
$w_{r0} = 82$ nm result in a high upper boundary $N_*$ for the
particle number of the Tonks gas. For the commonly used $^{87}$Rb
isotope with a scattering length $a=5$ nm one finds $N_* = 420$.
Although this is still a fairly low value it should be possible to
experimentally realize a small $^{87}$Rb Tonks gas. However, more
promising would be the use of the $^{85}$Rb isotope, where a
Feshbach resonance can be used to tune the normally negative
scattering length to positive values of several hundred nanometers
magnitude \cite{CornishCRCW00}. As $N_*$ is proportional to the
square of the scattering length even a moderate increase to $a =
50$ nm would make it possible to create a larger Tonks gas, with
say $N = 2000$ atoms, which should be easily detectable.

The Bessel beam trap offers some advantages compared to
alternative suggested approaches using magnetic waveguides
\cite{Key,Dekker,Thy} and a hybrid magnetic-optical trap
\cite{BonBurDet00}. Firstly, it involves only a very simple {\it
all-optical} system for which the aspect ratio of the trap may be
controlled simply by varying the Gaussian spot size incident on
the axicon. More specifically, being all-optical, it does not
involve material surfaces as in magnetic waveguides, which can
cause matter-wave decoherence \cite{HenWil99,HenPotWil99}.
Furthermore, it allows for the possibility of trapping multiple
magnetic sublevels and the investigation of multi-component Tonks
gases, which would not be possible in the hybrid magnetic-optical
trap of Bongs {\it et al.} \cite{BonBurDet00}.
\section{Summary and conclusions}
In this paper we have given an overview of our work on the quantum
dynamics of 1D trapped Tonks gases. In particular, we hope to have
shown that Tonks gases display a rich variety of behaviors that
are worthy of experimental investigation. The Fermi-Bose mapping
approach to solving for the 1D Tonks shows that many features are
shared between the Tonks gas of impenetrable bosons and a 1D gas
of non-interacting fermions: they have the same ground state
energies and density profiles, and neither shows true BEC.
However, the Tonks and Fermi gases differ markedly in their
momentum spectra, and we have devised a scheme to measure this
difference \cite{momentum}. Furthermore, we have shown that the 1D
Tonks gas can show dark soliton-like solutions which are typically
associated with solutions of the mean-field Gross-Pitaevskii
equation. This is so in spite of the fact that mean-field theory
greatly overestimates the coherence properties of the Tonks gas,
as we demonstrated in our simulations of split and recombined
Tonks gases.

Our work presented here shows that Tonks gases have some degree of
coherence but much less than a true BEC, and this may have
implications for proposed atomic interferometers employing high
density, eg. atom laser, sources. In particular, strong transverse
confinement is desirable in atomic interferometers to avoid
multi-transverse mode effects, and at the same time one would like to
keep the atomic density down to avoid many-body shifts, but these
are exactly the requirements for the Tonks limit! Therefore it is very
important to study the BEC-Tonks transition \cite{DunLorOls01} and
its effect on the performance of atom interferometers. We hope
that our work is a first step in this direction.
\vspace{0.2cm}

\noindent This work was supported at the University of Arizona by
Office of Naval Research grant N00014-99-1-0806 and also the US
Army Research office. We thank Joe Triscari for collaborations on
the ground state properties of the Tonks gas, and the work on
Bessel optical dipole traps was conducted in collaboration with J.
Arlt and K. Dholakia from St. Andrews University, Scotland, and J.
Soneson at the University of Arizona.
\begin{figure}
\caption{Gray-scale plots of the dimensionless pair distribution function
$x_{osc}^2\cdot D(Q_1,Q_2)$ as a function of the dimensionless
coordinates $Q_1$ and $Q_2$, for a) $N=2$, b) $N=6$, and c) $N=10$.}
\label{Fig:one}
\end{figure}
\begin{figure}
\caption{Gray-scale plots of the dimensionless reduced
density matrix $x_{osc}\cdot\rho_1(Q,Q')$ as a function of the
dimensionless
coordinates $Q$ and $Q'$, for a) $N=2$, b) $N=6$, and c) $N=10$.}
\label{Fig:two}
\end{figure}
\begin{figure}
\caption{Occupation of the natural orbitals: a) fraction of atoms
in the lowest orbital $f_0=\lambda_0/N$ versus $N$, and b)
$\lambda_j$ versus orbital number $j$ for
$N=2$ (circles), $N=6$ (stars), and $N=10$ (squares).}
\label{Fig:three}
\end{figure}
\begin{figure}
\caption{Dimensionless momentum distribution $k_{osc}\cdot n(\kappa)$ versus
normalized momentum $\kappa=k/k_{osc}$ for $a) N=2$, b) $N=6$, and
c) $N=10$.}
\label{Fig:four}
\end{figure}
\begin{figure}
\caption{Angular cross section versus angle
$\sin(\theta)\approx\theta$ for $N=10$. The dashed line is for the
1D gas of impenetrable bosons and the solid line is for the
corresponding system of non-interacting fermions} \label{Fig:five}
\end{figure}
\begin{figure}
\caption{Scaled density $\rho(x,t)L$ versus scaled position around
the ring $x/L$ for $N=51$, $t/\tau_e=0.051$, and
$\Delta\theta=\pi$ (solid line), and $\Delta\theta=0.5\pi$ (dashed
line). Due to symmetry we show only half of the ring $-L/2<x<0$,
the phase-jump being imposed at $x=-L/4$.} \label{Fig:six}
\end{figure}
\begin{figure}
\caption{Dark soliton velocity $|v|/c$ scaled to the speed of
sound $c$ as a function of phase-shift $\Delta\theta/\pi$ for
$N=51$.} \label{Fig:seven}
\end{figure}
\begin{figure}
\caption{Exact many-body theory simulation of the cool, cut,
interfere scenario. The figure shows a gray-scale plot of the
particle density $\rho_N(x,t)$ as a function of $\omega t$
(horizontal axis) and position $x/x_0$ (vertical axis), with white
being the highest density, for $N=10$, $V_B=20\hbar\omega$,
$w=3x_0$, and $\omega t_{pot}=3$.} \label{Fig:eight}
\end{figure}
\begin{figure}
\caption{Mean-field theory simulation of the cool, cut, interfere
scenario. The figure shows a gray-scale plot of the particle
density $\rho(x,t)=|\Phi(x,t)|^2$ as a function of $\omega t$
(horizonatal axis) and position $x/x_0$ (vertical axis).
Parameters are the same as Fig. 7.} \label{Fig:nine}
\end{figure}
\begin{figure}
\caption{Illuminating an axicon with a LG mode of order $\ell$
produces a Bessel beam of the same order within the shaded
region.} \label{Fig:ten}
\end{figure}
\newpage
\includegraphics*[width=0.4\columnwidth]{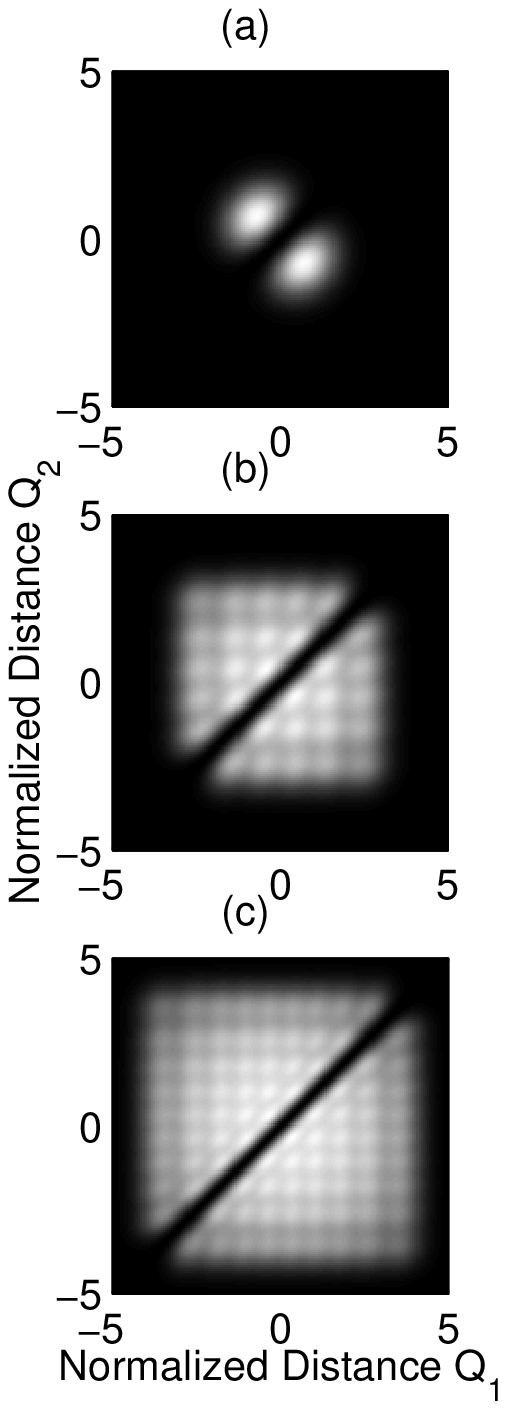}
\begin{center}
{\bf Figure 1}
\end{center}
\newpage
\includegraphics*[width=0.4\columnwidth]{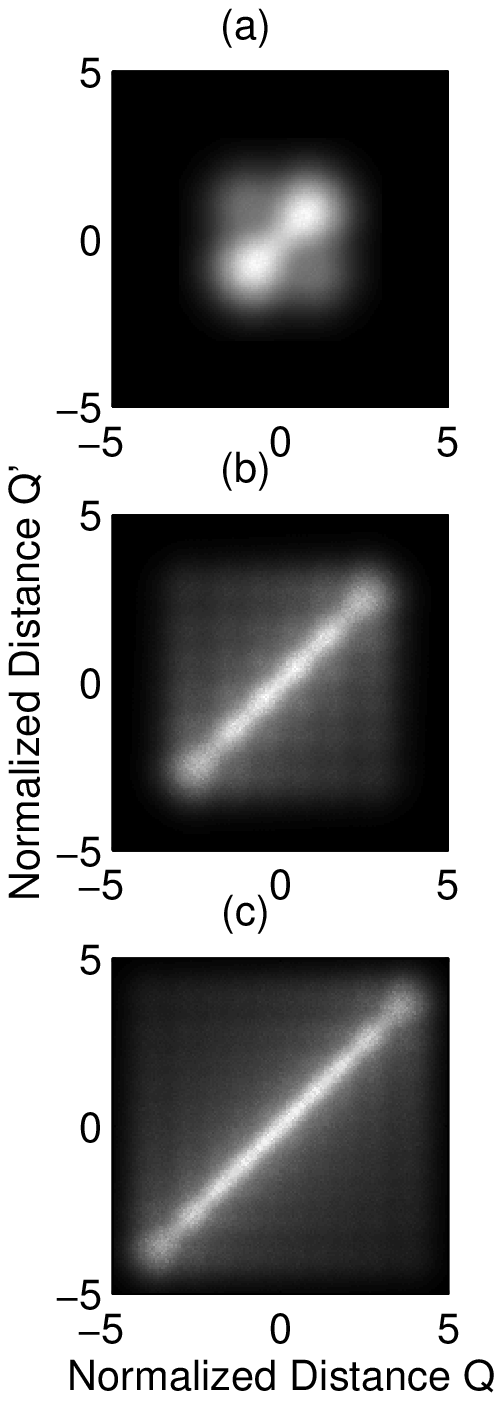}
\begin{center}
{\bf Figure 2}
\end{center}
\newpage
\includegraphics*[width=1.0\columnwidth]{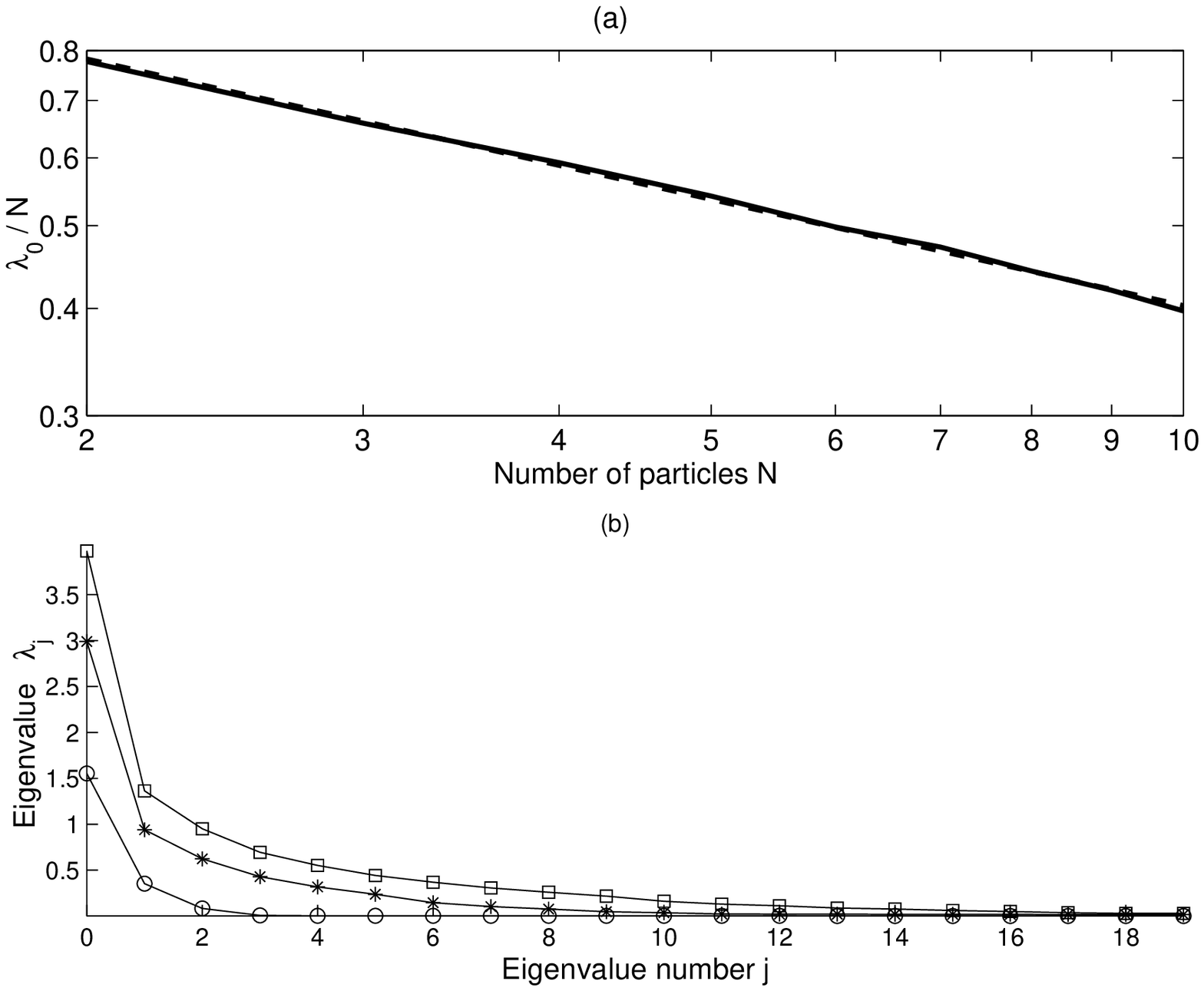}
\begin{center}
{\bf Figure 3}
\end{center}
\newpage
\includegraphics*[width=1.0\columnwidth]{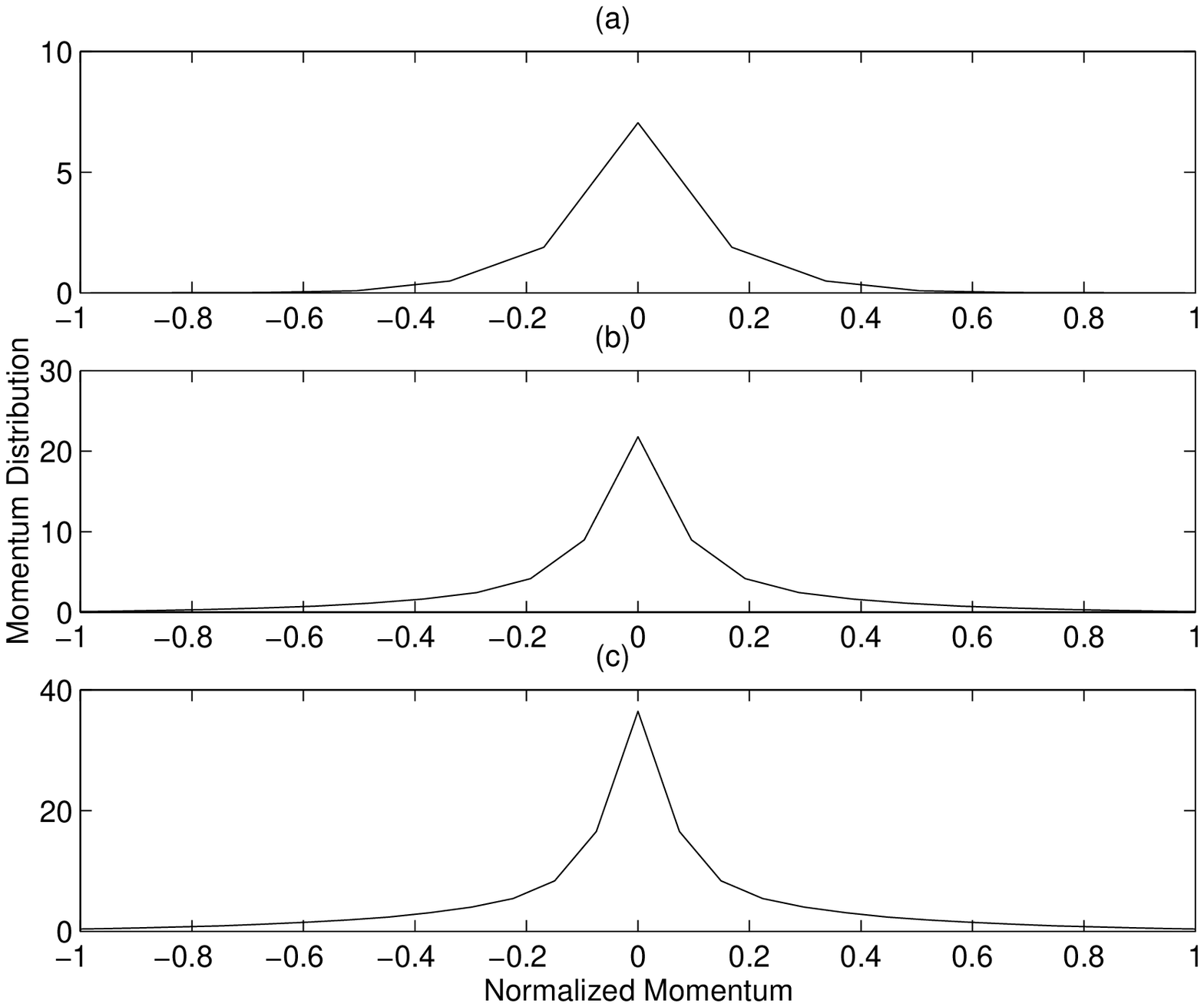}
\begin{center}
{\bf Figure 4}
\end{center}
\newpage
\includegraphics*[width=1.0\columnwidth]{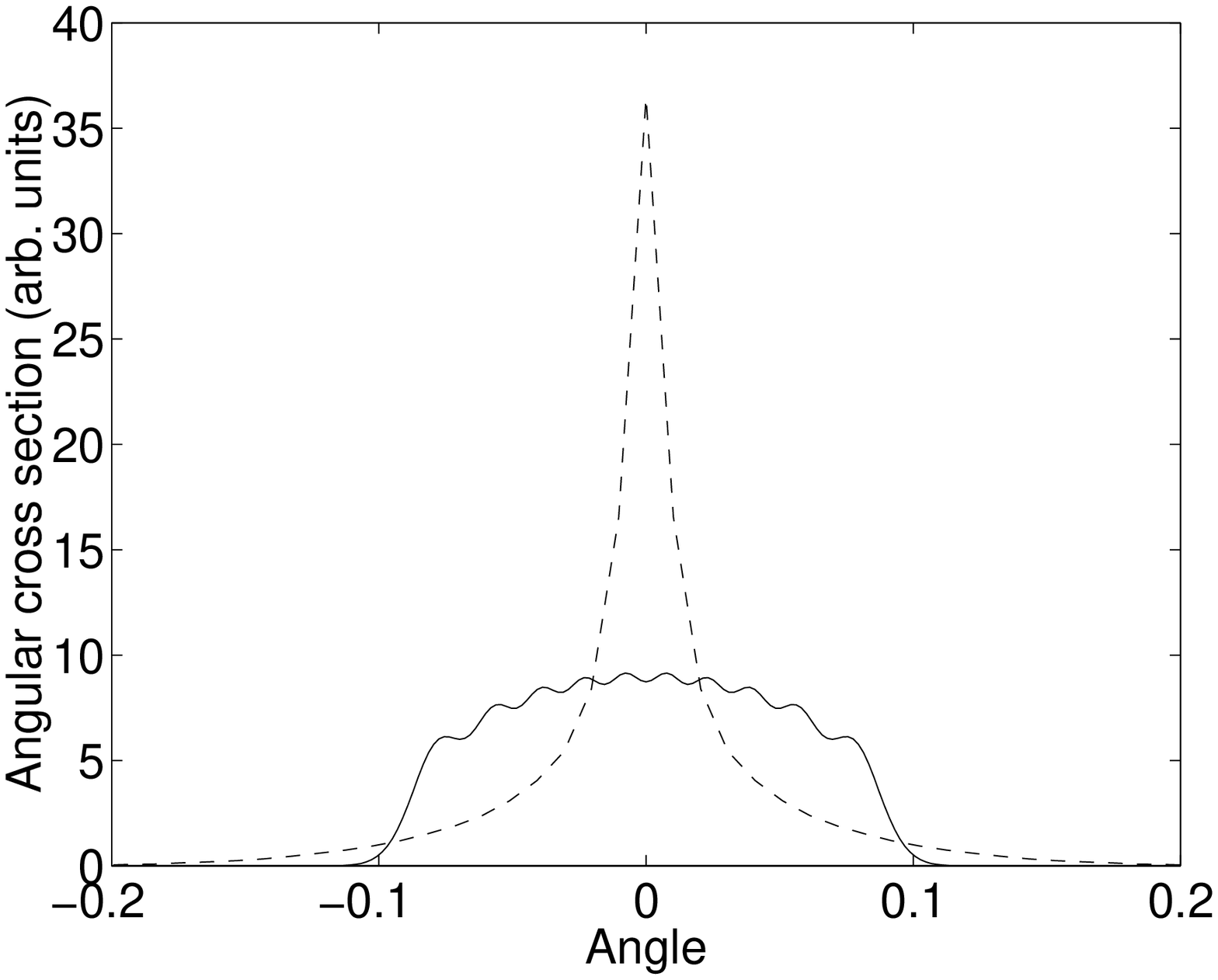}
\begin{center}
{\bf Figure 5}
\end{center}
\newpage
\includegraphics*[width=1.0\columnwidth]{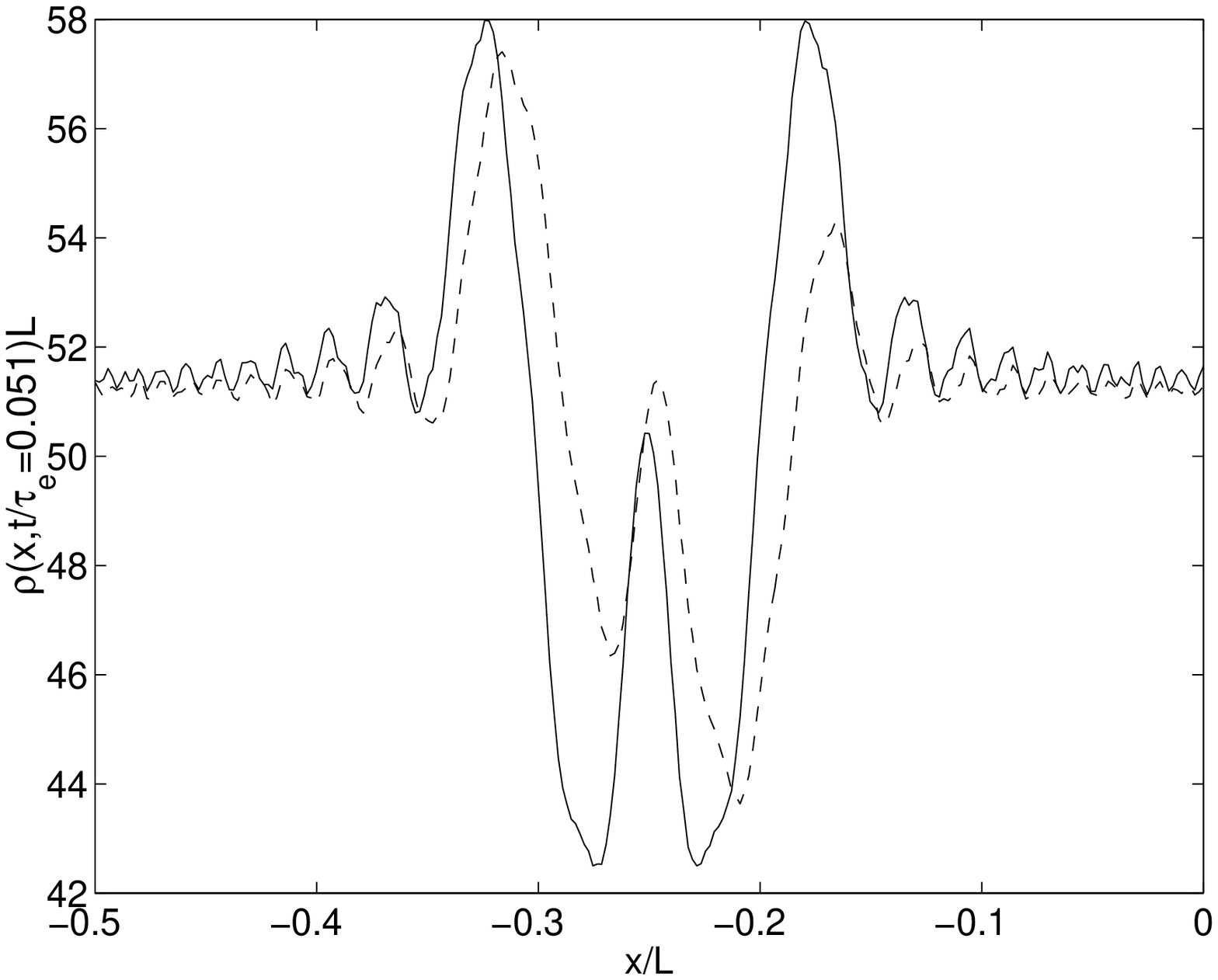}
\begin{center}
{\bf Figure 6}
\end{center}
\newpage
\includegraphics*[width=1.0\columnwidth]{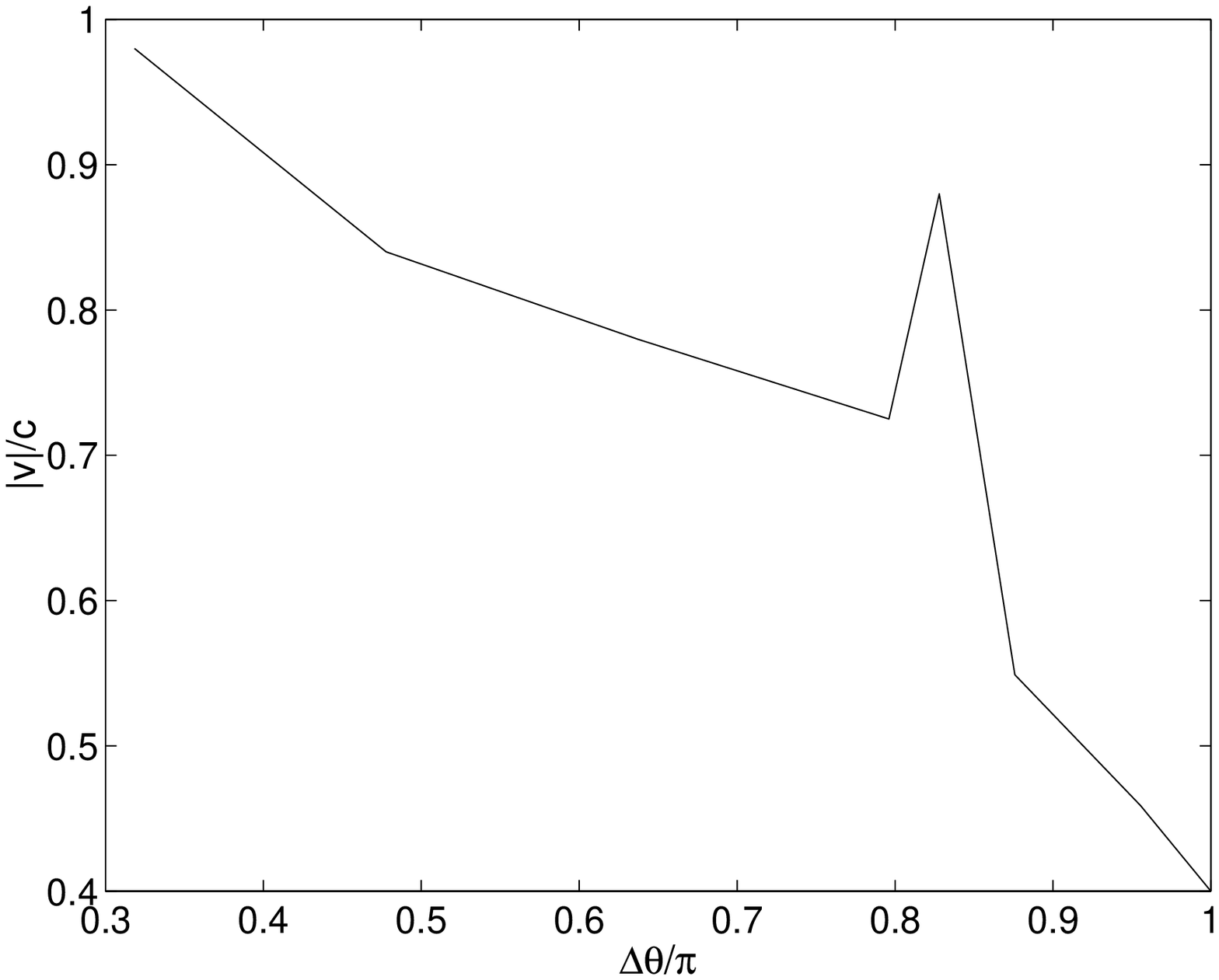}
\begin{center}
{\bf Figure 7}
\end{center}
\newpage
\includegraphics*[width=1.0\columnwidth]{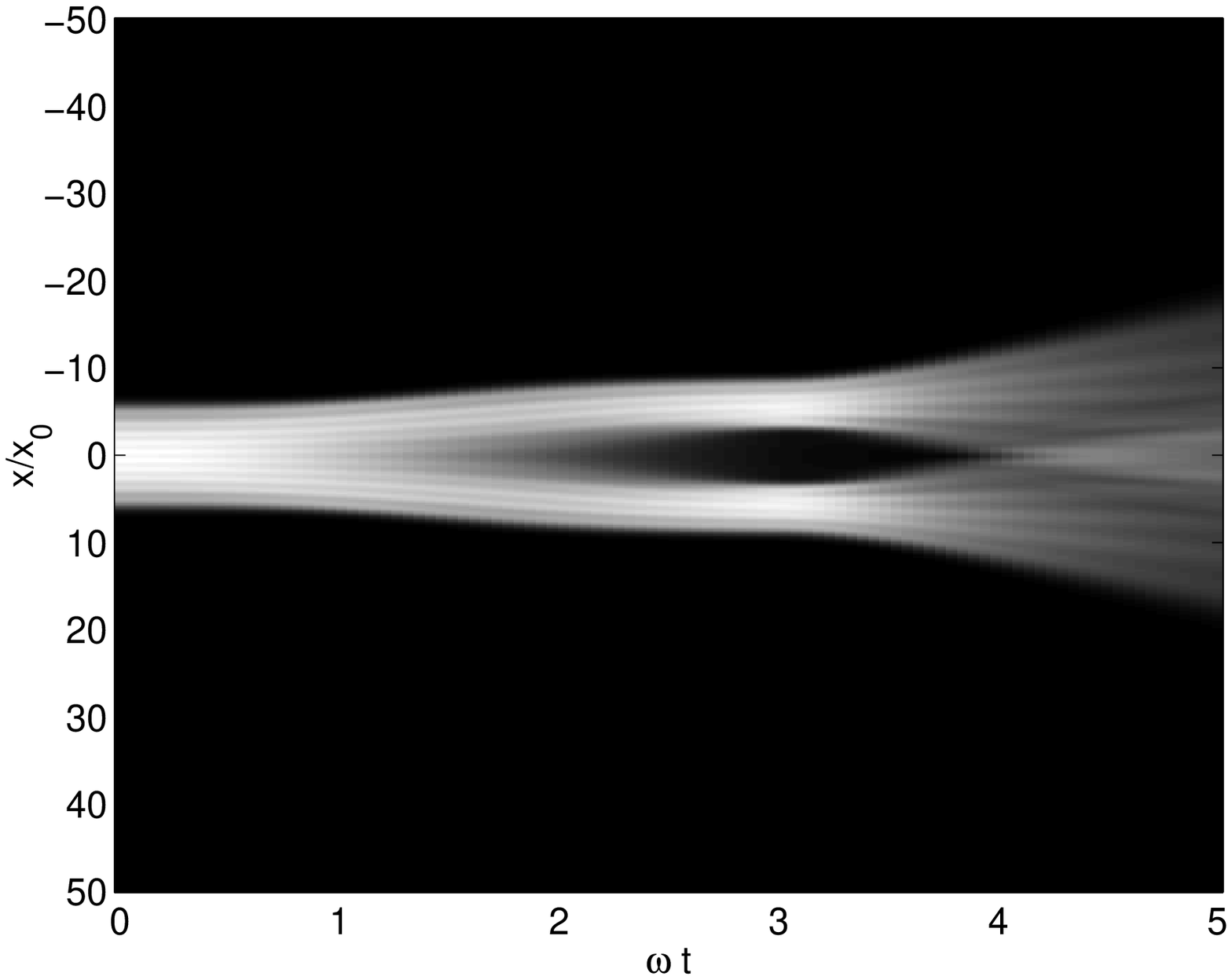}
\begin{center}
{\bf Figure 8}
\end{center}
\newpage
\includegraphics*[width=1.0\columnwidth]{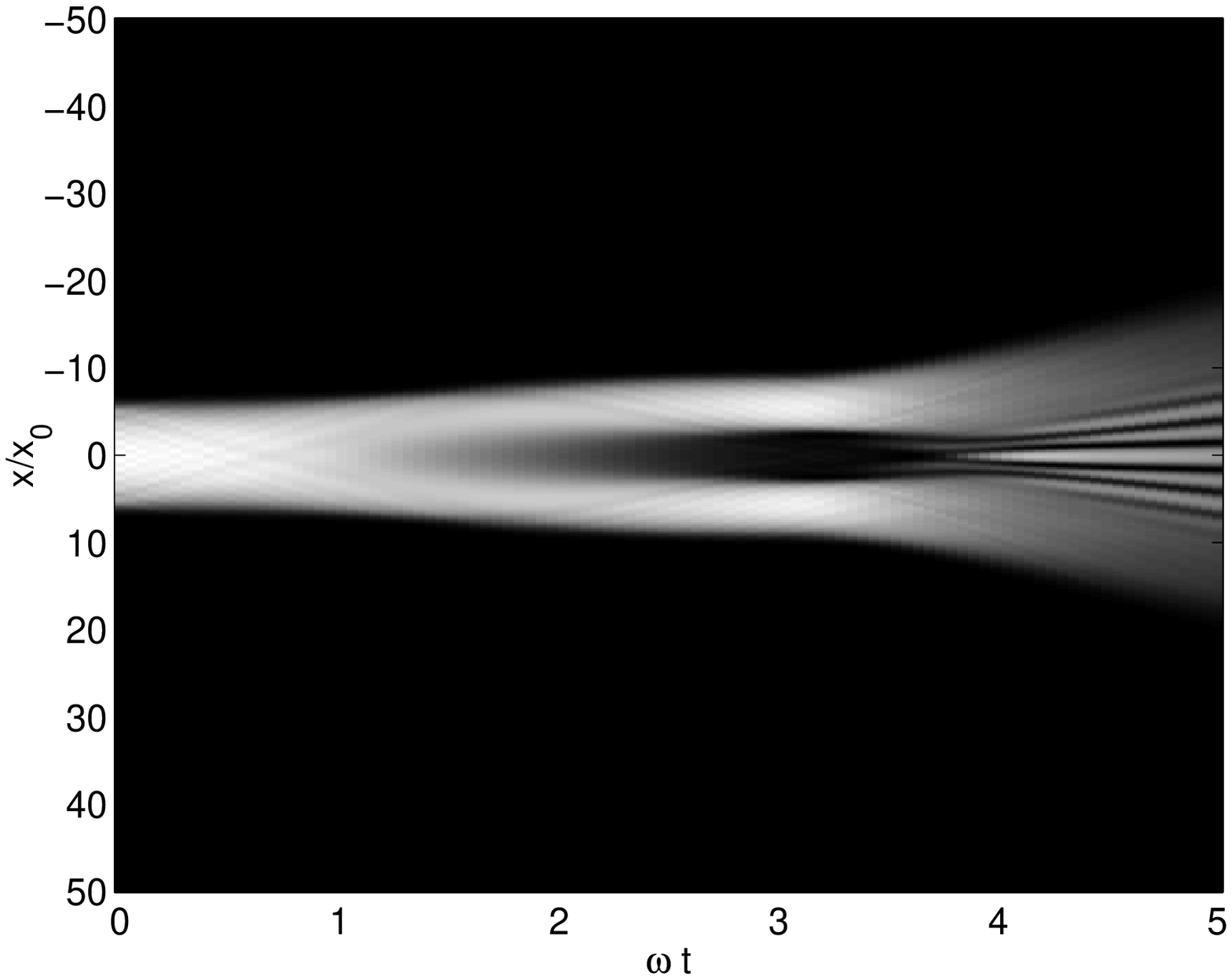}
\begin{center}
{\bf Figure 9}
\end{center}
\newpage
\includegraphics*[width=1.0\columnwidth]{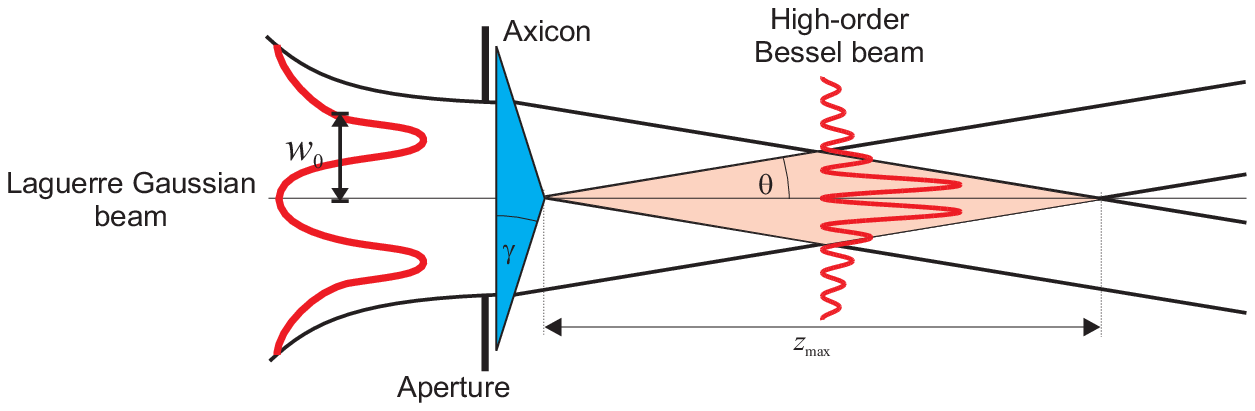}
\begin{center}
{\bf Figure 10}
\end{center}
\end{document}